# Vertical transport in InAs/GaSb superlattice: role of ionized impurity and interface roughness scatterings

**S. Safa [a], A. Asgari [1, a, b]**

[a]Research Institute for Applied Physics and Astronomy, University of Tabriz,

Tabriz 51665-163, Iran

[b]School of Electrical, Electronic and Computer Engineering,

The University of Western Australia, Crawley, WA 6009, Australia

**Abstract:**

We report on the vertical electron mobility versus temperature by applying the interface roughness scattering and ionized impurity scattering in InAs/GaSb superlattices. Using the Finite difference K.P method, we calculated the band structure of InAs/Gasb superlattices and then studied the transport properties of these systems. Several structural parameters such as layer thicknesses, interface roughness height, correlation length and ion density have been investigated and characterized that how the vertical mobility change with varying these parameters. Theoretical modeling results show that these two scattering mechanism are important in lower temperatures and thin layer systems.

**Introduction**

The tunability of the band gap in InAs/GaSb superlattices has made these systems technologically interesting for infrared lasers and detectors [1]. Moreover, superlattices (SLs) allow adjusting of the band structure for the noise reduction at higher temperatures [2]. This makes SLs useful for room temperature detector operation.

While the electrical properties and device performance are critically dependent on the transport properties of the system, they have not been investigated in detail yet.

The effects of interface roughness scattering in type II superlattices have been studied by Szmulowicz et al. [3-5] but, less attention has been paid to the effects of other scattering mechanisms on the vertical transport properties of electrons in InAs/GaSb SLs. Thus, the purpose of this paper is to report on a new technique for calculating the vertical electron mobility in type II superlattices based on a band structure and wave function obtain from K.P Finite Difference method, due to interface roughness and ionized impurity scattering.

---

[1] Corresponding Author: asgari@tabrizu.ac.ir, Tel: +98-411-3393005, Fax: +98-411-3347050.

## 2. Analytical Formalism

### 2.1. Electronic structure

For this study, we have solved the 8-band K.P Hamiltonian which is based on the Kane's formulation, using numerical Finite Difference method. The K.P Hamiltonian has the form of

$$H = \frac{p^2}{2m_0} + \frac{\hbar}{m_0}K.P + \frac{\hbar^2 k^2}{2m_0} + V + \frac{\hbar}{4m_0^2 c^2}(\sigma \times \nabla V) \cdot (\hbar K + P) \tag{1}$$

and

$$H \psi_{n,k} = E_{n,k} \psi_{n,k} \tag{2}$$

This full Hamiltonian will generate the matrix elements $H_{mn} = \langle u_{m,0} | H | u_{n,0} \rangle$, using the basis functions $u_{n,0} \uparrow$ and $u_{n,0} \downarrow$, which they are the eigen functions of Hamiltonian $H_0$, with spin added. The functions with $n \in \{s, p_x, p_y, p_z\}$ and

$$\psi_{n,k} = \sum_m C_{n,m} \uparrow u_{m,0} \uparrow + \sum_m C_{n,m} \downarrow u_{m,0} \downarrow \tag{3}$$

then $k_z$ is replaced by the differential operator $-i\frac{\partial}{\partial z}$ in the Hamiltonian and the Finite Difference method applied. [6, 7, 8] Accordingly we have calculated the electronic band structure of InAs/GaSb type-II SLs.

### 2.2. Scattering mechanisms

#### 2.2.1. Boltzmann equation

In the steady state condition and with applying a relaxation time approximation, the Boltzmann equation takes the form [9, 10]

$$\left.\frac{\partial f}{\partial t}\right|_c = eF \cdot \frac{1}{\hbar}\frac{\partial f}{\partial \mathbf{k}}, \tag{4}$$

where F is the electric field that counterbalances electron scattering. The scattering term for a system with volume V, is given by the integral

$$\left.\frac{\partial f}{\partial t}\right|_c = -\frac{V}{(2\pi)^3}\int\left[\Gamma(\mathbf{k},\mathbf{k}')f(\mathbf{k})(1-f(\mathbf{k}')) - \Gamma(\mathbf{k}',\mathbf{k})f(\mathbf{k}')(1-f(\mathbf{k}))\right]d\mathbf{k}', \tag{5}$$

and

$$f = f_0 + g, \qquad g(\mathbf{k}) = \tau(\mathbf{k})\frac{e}{\hbar}\left(F \cdot \frac{\partial E}{\partial \mathbf{k}}\right)\frac{\partial f_0}{\partial E} \qquad (6)$$

where $f_0$ is the equilibrium Fermi-Dirac distribution function and $g$ is the deviation from equilibrium, equation (5) contain both in-scattering and out-scattering terms. $\tau(\mathbf{k})$ is the relaxation rate and $\Gamma(\mathbf{k},\mathbf{k}')$ is the probability of scattering from initial state $\mathbf{k}$ to final state $\mathbf{k}'$. By applying the reversibility condition, the relaxation rates can be obtained as [11]

$$\frac{1}{\tau(\mathbf{k})} = \frac{V}{(2\pi)^3}\int d\mathbf{k}'\, \Gamma(\mathbf{k},\mathbf{k}')\frac{(1-f_0(\mathbf{k}'))}{(1-f_0(\mathbf{k}))} \times \left[1 - \frac{f_0(\mathbf{k})(1-f_0(\mathbf{k}))}{f_0(\mathbf{k}')(1-f_0(\mathbf{k}'))}\frac{g(\mathbf{k}')}{g(\mathbf{k})}\right] \qquad (7)$$

for vertical relaxation rates, we have

$$\frac{1}{\tau_\perp(k_r,k_z)} = \frac{V}{(2\pi)^3}\int \Gamma(k_r,k_z,k_r',k_z',\theta)k_r'dk_r'dk_z'd\theta \qquad (8)$$

where $\theta$ is the angle between $\mathbf{k}$ and $\mathbf{k}'$ [13].

### 2.2.2. Interface roughness scattering

In this section we study the interface roughness scattering in InAs/GaSb superlattices and its effects on electron mobility in these structures. The local atomic interface actually has a random variation which coupled the interface potential $V_e$, caused the perturbation potential as

$$\delta V(r_\parallel, z) = V_e \Delta(r_\parallel)\left[\delta(z+a) - \delta(z-a)\right] \qquad (9)$$

where $z = \pm a$ is the location of interfaces and $\Delta(r_\parallel)$ is assumed to be a random function at the in-plane position $(r_\parallel)$, which is usually taken to have a Gaussian correlation function with a characteristic roughness height of $\Delta$, and a correlation length $\Lambda$. This represents a length scale for roughness fluctuations along the interface, such that [12]

$$\langle \Delta(r_\parallel)\Delta(r_\parallel')\rangle = \Delta^2 e^{-\frac{|r_\parallel - r_\parallel'|^2}{\Lambda^2}} \qquad (10)$$

According to Fermi's golden rule, the scattering rate is given by [9,10]

$$\Gamma(\mathbf{k},\mathbf{k}') = \frac{2\pi}{\hbar}|M(\mathbf{k},\mathbf{k}')|^2 \delta(E(\mathbf{k})-E(\mathbf{k}')), \tag{11}$$

Where

$$|M(\mathbf{k},\mathbf{k}')|^2 = |\langle \mathbf{k}_\parallel, k_z |\delta V(r_\parallel)|\mathbf{k}'_\parallel, k'_z\rangle|^2. \tag{12}$$

Substituting Eq.12 intoEq.8, and carrying out the integration, we obtain the relaxation rates given by

$$\frac{1}{\tau_\perp(k_r,k_z)} = \frac{lV_e^2 m_\parallel}{2\hbar^3}\Delta^2\Lambda^2|\varphi_{k_z}(a)|^2$$

$$\int\left\{\exp\left(-\frac{|k_r^2+k_r'^2-2k_r k_r'\cos\theta|\Lambda^2}{4}\right)|\varphi_{k'_z}(a)|^2 \times \frac{1}{k_r}\times\delta(k_r-k'_r)\delta(E_{k_z}-E_{k'_z})\right\}k'_r dk'_r dk_z d\theta$$

(13)

where $k_r = |\mathbf{k}_\parallel|$, and $l = L_{InAs} + L_{GaSb}$ and $|\varphi_{k_z}(a)|$ is the value of wave function in the interface.

These equations given the relaxation time for electrons between two scatterings and they contains the effective masses and energy and the value of wave function in the interfaces. [13]

Screening effects have also been considered in our analysis, where the screening factor, which is multiplied by the transition matrix, can be written as [14, 15]

$$S_c(q) = \left(\frac{|\mathbf{k}_\parallel - \mathbf{k}'_\parallel|}{|\mathbf{k}_\parallel - \mathbf{k}'_\parallel|+q_s}\right)^2 = \left(\frac{\sqrt{k_r^2 - 2k_r k'_r \cos\theta + k_r'^2}}{\sqrt{k_r^2 - 2k_r k'_r \cos\theta + k_r'^2}+q_s}\right)^2 \tag{14}$$

where $q_s$ is the Thomas-Fermi screening wave vector.

*2.2.3. Ionized impurity Scattering*

To examine the scattering from ionized impurity we have to pay special attention to the long-range nature of this potential. Becauseinteractionsare assumed over all space, the integral diverges and a cut-off mechanism must be invoked to limit the integral. One approach is just to cut off the integration at the mean impurity spacing, the so-called Conwell-Weisskopf [16] approach. If one considers the faster potential drop and its screening length is the order of Debye screening length in non-degenerate materials, results in an integral which converges without further approximations (Brooks and Herring [17]).

In this paper we follow the second approach and the screened potential has been written below [18]

$$V(r) = \frac{e^2}{4\pi\varepsilon_\infty r}\exp(-q_d r), \tag{15}$$

where the Debye wave vector $q_d$ is the inverse of the screening length, and is given by

$$q_d^2 = \frac{ne^2}{\varepsilon_\infty k_B T} \tag{16}$$

Here $\varepsilon_\infty$ is the high-frequency permittivity, and n is the density of electrons in the conduction band.

$$\Gamma(\mathbf{k},\mathbf{k}') = \frac{2\pi}{\hbar}\left(\left|\left\langle \mathbf{k}_\parallel, k_z \left| V(r) \right| \mathbf{k}'_\parallel, k'_z \right\rangle\right|^2\right)\delta(E(\mathbf{k})-E(\mathbf{k}')) = \frac{2\pi}{\hbar}\left(\left|V(\mathbf{k}-\mathbf{k}')\right|^2\right)\delta(E(\mathbf{k})-E(\mathbf{k}')) \tag{17}$$

where $\psi_{\mathbf{k}_\parallel,k_z}(\mathbf{r}) \approx \left|\mathbf{k}_\parallel, k_z\right\rangle$ is the electron wave function, and for i'th ion we have,

$$V_i(\mathbf{k}'-\mathbf{k}) = \int \psi^*_{\mathbf{k}'_\parallel,k'_z}(r) V(r) \psi_{\mathbf{k}_\parallel,k_z}(r) dr \tag{18}$$

For $N_I$ ionized impurity, the scattering rate equation has the form of

$$\Gamma(\mathbf{k},\mathbf{k}') = \frac{2\pi N_I}{\hbar}\left(\left(\frac{Ze^2}{\varepsilon_\infty L^3}\right)^2 \frac{1}{\left|\mathbf{k}'-\mathbf{k}\right|^2+q_d^2}\int_0^\infty \exp(-i|\mathbf{k}'-\mathbf{k}|.z)\left|\varphi_{k_z}(z)\right|^2 dz\right)\delta(E(\mathbf{k})-E(\mathbf{k}')) \tag{19}$$

The vertical relaxation rates of electron is given by

$$\frac{1}{\tau_\perp(\mathbf{k})} = \frac{L^3}{(2\pi)^3}\int_0^\infty \Gamma(\mathbf{k},\mathbf{k}')d\mathbf{k}' \tag{20}$$

### 2.2.4. Electron Mobility

Hence, the vertical mobilities can be found from

$$\mu_\perp = -e\frac{\int \tau_\perp(\mathbf{k}) v_z^2 \left(\frac{\partial f_0}{\partial E}\right)d\mathbf{k}}{\int f_0(\mathbf{k})d\mathbf{k}} \tag{21}$$

where

$$v(\mathbf{k}) = \frac{1}{\hbar}\frac{\partial E}{\partial \mathbf{k}} \tag{22}$$

$$E(\mathbf{k}) = E(k_\parallel) + E_{k_z} = \frac{\hbar^2 k_\parallel^2}{2m_\parallel} + E_{k_z} \tag{23}$$

## 3. Results and Discussion

To model the transport properties of the electrons in the InAs/GaSb superlattices, we have computed the band structure for 7 structures with layer widths $l\,nm\,InAs/l'nm\,GaSb$ that $l,l' = 2,4,6,8$ that all infrared spans included. Knowing the energies and wave function for each structure, we have calculated the vertical scattering rates for electrons via two dominant scattering mechanisms.

For interface roughness scattering, we study the effect of three effective parameters that they are $T, \Lambda, \Delta$. Although the interface roughness scattering is temperature independent, but the carrier distribution function is temperature dependent, so the electron mobility via interface roughness scattering will be temperature dependent slightly, as depicted in equation (21). Figures 1-4 show the calculated vertical mobilities limited by interface roughness scattering as a function of temperature. In figures 1 & 2 we show the electron mobility for fixed 4 nm InAs/2 nmGaSb superlattice as a function of temperature for different $\Lambda$ and $\Delta$. Results in Fig. 1 shows that for low temperature the mobility rises, and it is because of the value of $\frac{\partial f_0}{\partial E}$, is a ascending function of temperature and in lower temperature the denominator of equation 21 is nearly constant. Also for higher temperature the electron density becomes larger and the mobilities decrease smoothly.

As can be seen clearly for smaller $\Lambda$, the mobility is high and it is dropping rapidly by increasing the correlation length of roughness as far as reaching a saturation value. For smaller correlation length, the mobility reaches the maximum value at temperature about 50 K, and this maximum point moves toward higher temperature for larger values of $\Lambda$, this is because of dependence of mobility to correlation length as $\left(\mu^{-1} \propto \Lambda^2 \exp\left(-\frac{\Lambda^2}{4}\right)\right)$.

In Fig. 2 the vertical mobility has been investigated as a function of temperature and roughness height. As mentioned above, the mobility variation versus temperature, caused by temperature dependence of distribution function. Also the results show that the mobility decreases with increasing the roughness height, and the mobility is proportional to $\Delta^{-2}$.

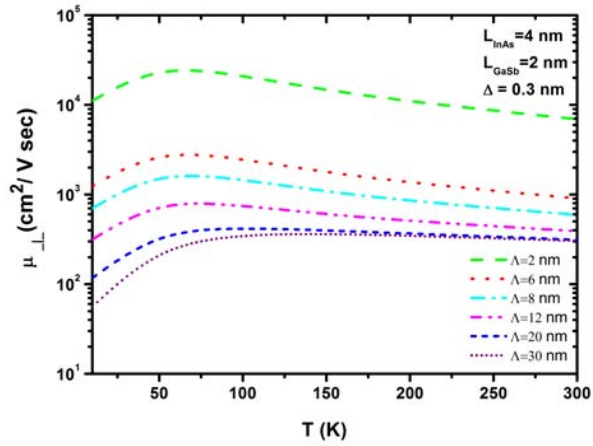

Fig.1. Calculated screened vertical mobility for electrons in a strained (40/20) InAs/GaSb SL vs temperature and correlation length. (Only the interface roughness scattering applied)

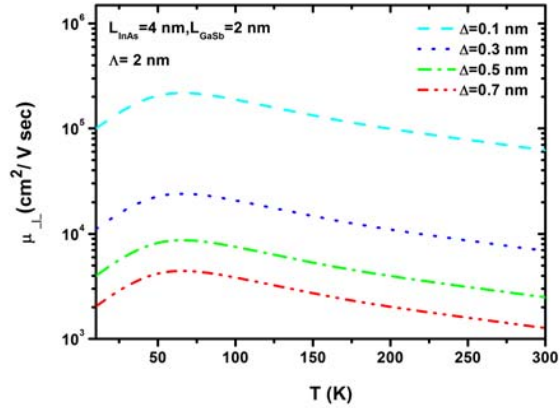

Fig.2. Calculated screened vertical mobility for electrons in a strained (40/20) InAs/GaSb SL vs temperature and roughness height. (Only the interface roughness scattering applied).

In figures 3&4 we study the temperature dependence vertical mobility, also the layer width in superlattices is the variable parameter. As previously mentioned, the temperature dependent parameter in this case is the distribution function of carriers, and this function explains the increase and decrease in mobility versus temperature. The calculated vertical mobility of the InAs/GaSb superlattices with different well width is shown in Fig. 3, as can be seen the mobility increases with increasing width of the InAs well, which follows the Gold model [19]. This behavior is because of reduction in electron effective masses and energy with increasing the well width. In 8nmInAs/2nmGaSb system in 0-50 $^0$K, the mobility is becomes about ten times larger, and it is because of very low scattering rate and carrier density in this range. By increasing the temperature the electron density increase and the mobility becomes smaller again.

We have presented the vertical mobility under various barrier widths in Fig. 4, a remarkable characteristic of the diagram is that we have seen the large mobility at higher temperature for 4nm/4nm superlattice that it is because of the optimum effective masse and energy and wave function of electron in these conditions, which cause the lower scattering rate for this structure. Also by increasing the barrier width the tunneling of electrons between the layers decreasing and also the slob of conduction band become smaller, so in equation 21, we have smaller $v_z$, as a result smaller mobility.

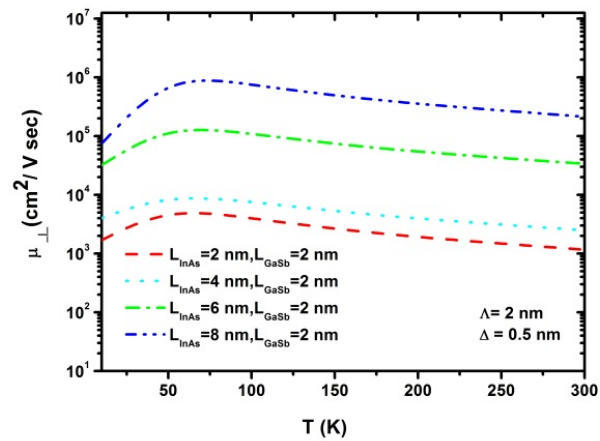

Fig.3.Calculated screened vertical mobility for electrons

vs temperature and well width.(Only the interface roughness scattering applied).

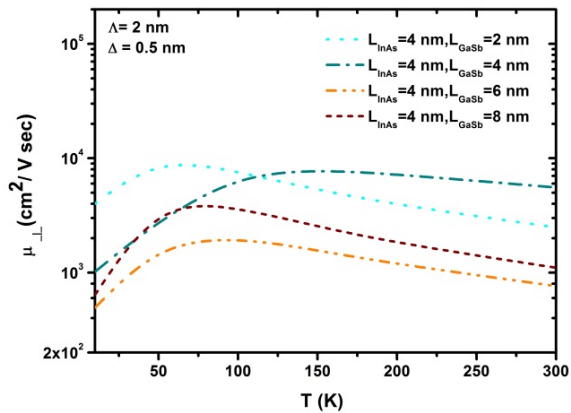

Fig.4.Calculated screened vertical mobility for electrons

vs temperature and barrier width.(Only the interface roughness scattering applied).

Using equation 20 we have calculated the vertical mobility via ionized impurity scattering. In Fig. 5 we have plotted the calculated electron mobilities as a function of the temperature and impurity density, unlike the previous case, the ionized impurity scattering rate is a function of temperature. By increasing the temperature the electron density (n) increase exponentially and in cause growth in screening length (q). Since the scattering rates inversely proportional to square screening length, the mobility sharply rises with increasing the temperature. In the equation 19 we assume that the scattering rate is directly proportional to number of impurity and in Fig. 5 we showed the numerical results. In Fig. 6 we studied the effect of temperature on electron mobility for different structures. As mentioned before, the mobility increase with increasing the temperature, also by increasing the well width, because the energies and effective mass of electrons and the values of wave function decrease, the scattering rate, (Eq. 19), drop and the mobility becomes larger.

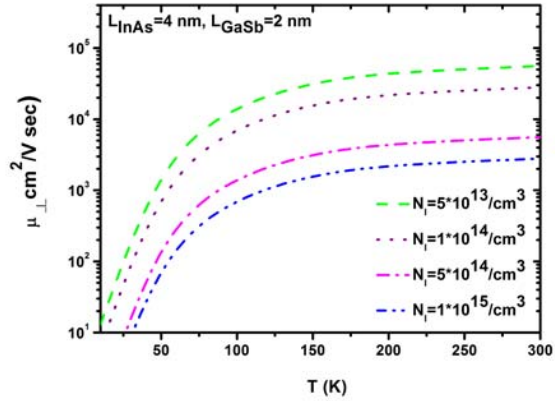

Fig.5.Calculated screened vertical mobility for electrons in a strained (40/20)InAs/GaSbSL

vs temperature and impurity density.(Only the ionized impurity scattering applied).

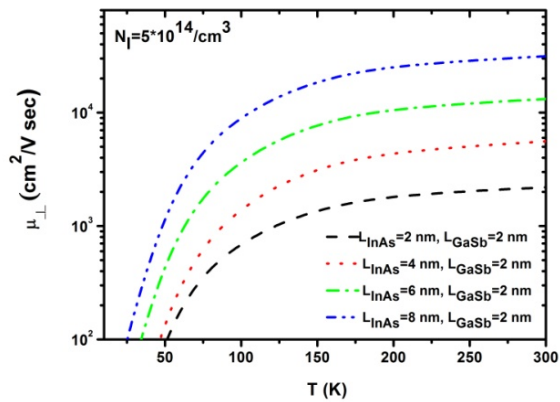

Fig.6.Calculated screened vertical mobility for electrons

vs temperature and well width.(Only the ionized impurity scattering applied).

## 4. Conclusions

We have calculated the vertical electron mobility due to interface roughness scattering and ionized impurity scattering in InAs/GaSb type-II superlattices using energy spectra and wave functions obtained via a K.P Finite Difference method. Using the Boltzmann equation and Fermi golden role we obtain the vertical scattering rates and then we study the effect of other parameters on electron mobility. The results presented in this paper indicate that increasing the temperature cause the rise in mobilities limited by ionized impurity scattering. Although the interface roughness scattering rate is temperature independent, the distribution function caused variation the mobilities by temperature.

Also we study the effects of some parameters on mobilities such as interface roughness interface roughness correlation length and roughness height and impurity density, found that they reduce the mobility. We repeated all calculation for different structures and the results show that, increasing the InAs width in superlattice, makes the lower energy and wave function value and increases the mobilities, and the 4nm InAs/4nm GaSb structure has an optimum mobility at the room temperature.